\documentclass[12pt]{article}
\usepackage{amssymb}
\usepackage{amsfonts}
\begin{document}

\baselineskip18truept
\begin{center}
{\LARGE{\scshape BF system} --- encyclopedic entry%
\footnote{
\emph{Concise Encyclopedia of Supersymmetry
And Noncommutative Structures in Mathematics and Physics},
Duplij, S.; Siegel, Warren; Bagger, Jonathan (Eds.)
2005, pages 53--54.}}
\end{center}
\begin{center}
{\Large\mdseries Bogus{\l}aw Broda%
\footnote{The work has been supported by the grant of the 
University of \L\'od\'z.}}
\end{center}
\begin{center}
\large\textit{
Department of Theoretical Physics\\
University of \L\'od\'z\\
Pomorska 149/153\\
PL 90--236 \L\'od\'z\\
Poland%
\footnote{E-mail: {\tt bobroda@uni.lodz.pl}}}
\end{center}
\bigskip

BF SYSTEM, a field-theory system consisting of the two 
fields $B$ and $F$. On the $D$-dimensional manifold ${\cal 
M}^D$ ($D \geq2$), the classical action of the 
(\emph{pure}) BF system is defined by the 
metric-independent expression
$$ 
\int_{{\cal M}^D}{\rm Tr}(B_{D-2}\wedge F_A),
\eqno(1)
$$
where $B_{D-2}$ is a Lie algebra-valued differential 
$D-2$-form, and $F_A$ is the curvature of the connection 
form $A$, $F_A=d A+{1\over2}[A,A]$. BF system constitutes 
an important class of topological gauge field theory models 
of the, so-called, Schwarz type, akin to Chern-Simons 
system [1] (further, older references therein). Gauge 
symmetry of BF system is of the two types: (1) standard 
Yang-Mills symmetry with infinitesimal transformations
$$
\delta_{\rm YM} A= d_A\omega, \qquad \delta_{\rm YM} 
B_{D-2}=[B_{D-2},\omega],
$$
and (2) $D-2$-form $B$ symmetry ($D \geq3$),
$$
\delta_{\rm B} A=0, \qquad \delta_{\rm B} B_{D-2}=d_A 
\Lambda_{D-3},
$$
where $\Lambda_{D-3}$ is a Lie algebra valued $D-3$-form, 
and $d_A=d+[A,\cdot\,]$. Since the classical equations of 
motion corresponding to (1) are
$$
F_A=0,\qquad d_A B_{D-2}=0,
$$
we have, for a sufficiently large $D$, a chain of on-shell 
reducible gauge symmetries
$$
\delta_{\rm B}\Lambda_k=d_A \Lambda_{k-1},\qquad 
k=1,2,\dots,D-3,
$$
invoking, at quantum level, the formalism of Batalin and 
Vilkovisky (see [2], also for the issue of UV finiteness).

In the Abelian case [1], a generalization of the classical 
action (1) is 
$$
\int_{{\cal M}^D} B_p \wedge dA_{D-p-1}.
$$
The partition function $Z_p({\cal M}^D)$ yields a power of 
the Ray-Singer torsion $T_D$
$$
Z_p({\cal M}^D)=
\cases{
T_D^{(-1)^p},	& for $D$ odd;\cr
1,				& for $D$ even.\cr}
$$
Observables (Wilson surfaces) give rise to the linking 
number $\ell$
$$
\biggl<\exp{\biggl(\int_{\Sigma_B^p}B\biggr)}\,
\exp{\biggl(\int_{\Sigma_A^{D-p-1}}A\biggr)}\biggr>
=\exp{\left[i \ell \bigl(\Sigma_B^p,
\Sigma_A^{D-p-1}\bigr)\right]}.
$$

For non-Abelian BF system, there is a relation to the 
Ray-Singer torsion, too, and to various known knot and link 
invariants in $D=3$, whereas in $D>3$, the observables are 
expected to probe higher-dimensional topology, but the 
knowledge has not stabilized yet [3].

Besides the pure BF system (1) one can consider the, 
so-called, BF system with a \emph{cosmological} term, an 
(exterior) product of $n$ ($np=D$) $B$ fields
$$
\int_{{\cal M}^D}
{\rm Tr}\left(B_p \wedge\dots\wedge B_p\right).
$$
3D BF system with a cosmological term is directly related 
to Chern-Simons theory---in particular, the partition 
function yields the Turaev-Viro invariant ${\rm TV}({\cal 
M}^3)$ of the 3D manifold ${\cal M}^3$. Further 
generalizations can include SUSY.

One of the most exciting applications of BF systems is 
gravity [4]. 3D gravity is directly identified with a BF 
system, and the cosmological term plays the role of the 
cosmological term in gravity. 4D BF system can be a 
starting point for a novel formulation of 4D gravity 
(direction pioneered by Pleba\'nski).

Analogously to the case of akin Chern-Simons theory, BF 
system can be used for gauge-invariant generation of mass 
for gauge fields. Yet another application consists in a 
first-order formulation of ordinary Yang-Mills system, 
where the Yang-Mills theory can be considered as a 
deformation of topological system.


\begin{thebibliography}{X}
\bibitem{}
D.B.~Birmingham, M.~Blau, M.~Rakowski, G.~Thompson, Phys.\ 
Rep.\ {\textbf{209}} (1991) 129;
\bibitem{}
O.~Piguet, S.P.~Sorella, Algebraic Renormalization, 
Springer 1995;
\bibitem{}
A.S.~Cattaneo, P.~Cotta-Ramusino, C.A.~Rossi, Lett.\ Math.\ 
Phys.\ {\textbf{2}} (2000) 301; 
\bibitem{}
J.C.~Baez, Lect.\ Notes Phys.\ {\textbf{543}} (2000) 25.
\end{thebibliography}
\end{document}